\documentclass[aps,prl,
 amsmath,amssymb,
 amsfonts,twocolumn,nofootinbib,10pt
]{revtex4-2}

\usepackage{graphicx}   
\usepackage{dcolumn}    
\usepackage{bm}         
\usepackage{physics}    
\usepackage{hyperref}   
\usepackage{xcolor}     
\usepackage{scalerel}
\usepackage{tikz}
\usepackage{ulem}
\usetikzlibrary{svg.path}

\newcommand{\prlsection}[1]{%
  \noindent\textit{#1}---\ \ignorespaces}

\definecolor{orcidlogocol}{HTML}{A6CE39}
\tikzset{
  orcidlogo/.pic={
    \fill[orcidlogocol] svg{M256,128c0,70.7-57.3,128-128,128C57.3,256,0,198.7,0,128C0,57.3,57.3,0,128,0C198.7,0,256,57.3,256,128z};
    \fill[white] svg{M86.3,186.2H70.9V79.1h15.4v48.4V186.2z}
                 svg{M108.9,79.1h41.6c39.6,0,57,28.3,57,53.6c0,27.5-21.5,53.6-56.8,53.6h-41.8V79.1z M124.3,172.4h24.5c34.9,0,42.9-26.5,42.9-39.7c0-21.5-13.7-39.7-43.7-39.7h-23.7V172.4z}
                 svg{M88.7,56.8c0,5.5-4.5,10.1-10.1,10.1c-5.6,0-10.1-4.6-10.1-10.1c0-5.6,4.5-10.1,10.1-10.1C84.2,46.7,88.7,51.3,88.7,56.8z};
  }
}

\newcommand\orcidicon[1]{\href{https://orcid.org/#1}{\mbox{\scalerel*{
\begin{tikzpicture}[yscale=-1,transform shape]
\pic{orcidlogo};
\end{tikzpicture}
}{|}}}}

\begin{document}

\title{Microwave-controlled interactions and stripe formation of static--field--shielded polar molecules}

\author{T. Arnone Cardinale\,\orcidicon{0009-0000-7916-5091}}
\email{tiziano.arnone\_cardinale@fysik.lu.se}
\author{M. Schubert\,\orcidicon{0009-0001-3095-2921}}
\author{S.M. Reimann\,\orcidicon{0000-0003-1869-9799}}
\affiliation{Mathematical Physics and NanoLund, LTH, Lund University, Box 118, 22100 Lund, Sweden}

\date{\today}

\begin{abstract}
We study polar molecules where short--range losses are suppressed by a shielding scheme involving a static electric field and an elliptically polarized microwave field. Using perturbation theory, we derive the effective interaction potential and validate it against coupled channel calculations. We identify a parameter regime where two--body losses are strongly suppressed and the extended mean-field description of dilute molecular Bose–Einstein condensates is justified. We calculate the collective excitations and show that intriguingly, supersolidity in quasi-two-dimensional confinement emerges as a stripe phase even at small values of microwave ellipticity.
\end{abstract}

\maketitle

The creation of Bose--Einstein condensates (BECs) of magnetic atoms~\cite{Griesmaier2005,Stuhler2005,Lu2011,Aikawa2012,Miyazawa2022} opened up new possibilities to explore quantum matter with long-range anisotropic interactions~\cite{Werner2005, Lahaye2009, Chomaz2023}. Exotic phases such as self-bound quantum droplets~\cite{Kadau2016,Ferrier2016,Schmitt2016} were found, in which the self-binding is caused by balancing the attractive part of the dipole-dipole interactions with repulsive quantum fluctuations~\cite{Lima2011,Waechtler2016QuantumFilaments,chomaz2016quantum}. Prominently, this led to the realization of the  supersolid state, where phase-coherence and crystalline order co-exist~\cite{Bottcher2019,tanzi2019observation,Chomaz2019}.

Compared with magnetic atoms, the electric dipole moments of polar molecules can be much larger, allowing to reach the strongly correlated limit~\cite{Buchler2007}.
Ultracold gases of polar molecules have been experimentally realized~\cite{Ni2008, takekoshi2014ultracold, molony2014creation,Park2015,guo2016creation,demarco2019degenerate,Schindewolf2022,stevenson2023ultracold}, also opening new perspectives for the quantum-control of chemical reactions~\cite{Ospelkaus2010,Idziaszek2010,Zhao2022}. 
Reaching the condensed state with bosonic polar molecules was long hampered by severe losses ~\cite{Mayle2013,Gregory2019,Bause2023,Schindewolf2026}. However, the losses can be mitigated by collisional shielding techniques~\cite{Will2016,Karman2018,Schindewolf2022,KarmanDouble,Quener2016,MukherjeeShielding,Gorshkov2011, HoTuning} which create a short-range barrier. In particular, double microwave shielding~\cite{KarmanDouble}  paved the way for the experimental realization of BECs of polar molecules~\cite{Bigagli2024,Shi2026bose}.
Equally important, shielding also provides precise control over the strength and anisotropy of dipolar interactions, enabling access to a rich variety of many-body phases~\cite{Schindewolf2026}, including superfluids \cite{Gorshkov2011,Li2025,DengEffective}, self-bound quantum droplets~\cite{Langen2025}, droplet arrays~\cite{Zhang2026} and self-bound monolayer crystals~\cite{Ciardi2025}. Recently, it was shown that mean-field theory can reliably predict the ground states of dipolar molecular gases in certain parameter regimes~\cite{Baena2025,ArnoneExploring}, enabling the study of large condensates, including the supersolid phase. While the effect of the dipolar interaction on the ground-state properties of molecular BECs has been extensively studied~\cite{Schindewolf2026}, the additional anisotropy introduced by certain shielding techniques~\cite{karman2019microwave,karman2020microwave,DengEffective,xu2025effective} remains largely unexplored.

In this Letter, we derive the effective interaction for polar molecules shielded by a static electric field and an elliptically polarized microwave field, as it has recently been suggested by Ho {\it et al.}~\cite{HoTuning} for a purely circular microwave. We study ground-state properties and dynamics of molecular BECs in a regime where collisional losses are suppressed and interactions are weak, making the use of the extended Gross-Pitaevskii approach viable. We identify the parameter space for this regime and demonstrate that the anisotropy of the effective interaction strongly modifies the ground state and the excitation spectrum of a large BEC. Intriguingly, even a small ellipticity of the microwave field induces a supersolid stripe phase in a quasi-two-dimensional molecular BEC. Unlike stripe phases known from atomic BECs with spin-orbit coupling~\cite{li2017stripe} or dipolar BECs with an external tilt of the dipole direction~\cite{Bombin2017dipolar,chandrashekara2026competing}, here the modulation is induced by the dressing field like the ones recently reported for dual-microwave-shielded molecules~\cite{Ciardi2026}. Our work establishes the connection between the well-known theory of collective excitations and two-dimensional supersolidity in dipolar molecular BECs.

\begin{figure}[t]
    \centering
    \includegraphics[width=\linewidth]{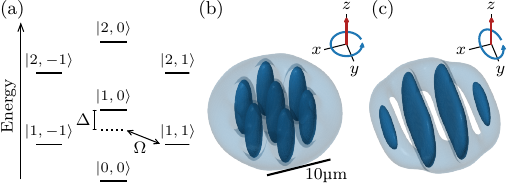}
    \caption{Supersolids of shielded polar molecules. (a) Energy level diagram of a molecule under an electric field and $\sigma^+$-polarized microwave. The microwave has Rabi frequency $\Omega$, detuning $\Delta$ and couples the $\ket{1,0}$ state to the $\ket{1,1}$ state. For imperfect polarization, the $\ket{1,-1}$ state is coupled as well. (b) Ground state of a gas of $3\times10^4$ CaF molecules under an electric field of $22.5~\mathrm{kV/cm}$ and microwave with $\Omega=60~\mathrm{MHz}$ and $\Delta=1.925\Omega$ in a pancake trap. The insets show the static electric field $\mathbf{F}$ (red) and the perpendicular, rotating microwave field $\mathbf{E}$ (blue). (c) Same as (b), but for an elliptically polarized microwave with $\xi=1^\circ$. The isosurfaces in (b) and (c) are shown for 5\% and 16\% of the peak density (with the same length scale as indicated in the figure).}
    \label{fig:schem}
\end{figure}

\indent \prlsection{Effective Potential}\label{sec:effective}
We consider the protocol proposed in \cite{HoTuning}, where a static electric field $\bm F$ aligned along $z$ provides shielding and an imperfectly circularly polarized microwave is used to control the dipolar interaction strength. The Hamiltonian of a single molecule in the presence of the static field is
\begin{equation}
    H_0 = B_\mathrm{rot}\bm J^2-\bm F\cdot\bm d,
\end{equation}
where $\bm J$ is the angular momentum operator, $\bm d$ is the dipole operator, and $B_\mathrm{rot}$ is the rotational constant of the molecule. The Hamiltonian $H_0$ has eigenstates $\ket{n,m}$, where $m$ is the projection of $\bm J$ along z. In this work, we restrict the Hilbert space to $n=0,1,2$ and consider a microwave of Rabi frequency $\Omega$ that couples the state $\ket{1,0}$ with $\ket{1,1}$ and $\ket{1,-1}$ with detuning $\Delta$. A schematic of the states involved is given in Fig.~\ref{fig:schem}(a). The electric field of the microwave is given by
\begin{align}
    \bm E(t) =  -\frac{E_0}{2}e^{-i\omega_\mathrm c t}(\bm e_+ \cos\xi  + \bm e_- \sin\xi ) +c.c.
\end{align}
with amplitude $E_0$, frequency $\omega_\mathrm c$, ellipticity $\xi$ and $\bm e_{\pm}=\mp(\bm e_x\pm i\bm e_y)/\sqrt 2$. The interaction of a single molecule with the field in the dipole approximation is $H_1 = -\bm d\cdot E(t)$. Moving to the rotating frame and taking the rotating wave approximation, the interaction becomes $H_1=\hbar\Omega(\ket{1,0}\bra{\xi_1} + \ket{\xi_1}\bra{1,0})/2$, where $\ket{\xi_1}=\cos\xi\ket{1,-1} +\sin\xi \ket{1,1}$ is the coupled state and the corresponding dark state is $\ket {\xi_2} = \sin\xi\ket{1,-1} -\cos\xi \ket{1,1}$. The Rabi frequency is given by $\Omega=E_0\langle1,0|\mathbf{d}|1,1\rangle/\hbar$.

In the dressed-state approach, the basis that diagonalizes the total Hamiltonian $H_0+H_1$ is $\mathcal B =\{\ket{0,0},\ket{+},\ket{-},\ket{\xi_2},\ket{2,m}\}$, with $\ket + = u\ket{1,0}+v\ket{\xi_1}$, $\ket - = v\ket{1,0}-u\ket{\xi_1}$, $u=\sqrt{(1+\Delta/\Omega_\mathrm{eff})/2}$, and $v=\sqrt{(1-\Delta/\Omega_\mathrm{eff})/2}$. The effective Rabi frequency is $\Omega_\mathrm{eff}=\sqrt{\Omega^2+\Delta^2}$. From this one-body basis we construct a basis of symmetrized two-molecule states, see {\it End Matter (Appendix~A)}. The dipole-dipole interaction between two molecules at a distance $r$ is given by
\begin{align}
    V_\mathrm{dd}(\bm r) = \frac{1}{4\pi\epsilon_0 r^3}\Big(\bm d_1\cdot \bm d_2 -3\frac{(\bm d_1\cdot {\bm r})(\bm d_2\cdot {\bm r})}{r^2} \Big).
\end{align}

We derive the effective interaction potential for molecules in the pair state $\ket{+}\ket{+}$ using perturbation theory~\cite{DengEffective,Mukherjee2025effective}. Such potential reads
\begin{equation}
    V_\mathrm{eff}(\bm r) = V_3(\bm r) + V_6(\bm r),
\end{equation}
where the first order term is the effective dipole-dipole interaction
\begin{equation}\label{eq:V3}
    V_3(\bm r) = \frac{C_3^{(1)}}{r^3}(1-3\cos^2\theta) + \frac{C_3^{(2)}}{r^3}\sin^2\theta\cos2\phi
\end{equation}
with coefficients
\begin{align}
    C_3^{(1)} &= \frac{(u^2d_{10,10}^{(z)}+v^2d_{11,11}^{(z)})^2 - (u\,v\,d_{10,11}^{(+)})^2}{4\pi\epsilon_0}\\
    C_3^{(2)} &= \frac{3(u\,v\,d_{10,11}^{(+)})^2\sin 2\xi}{4\pi\epsilon_0},
\end{align}
where $d_{\alpha,\beta}^{(j)}=\bra{\alpha}\bm d\cdot \bm e_{j} \ket{\beta}$ are the matrix elements of the dipole operator for each spherical component $j$.

The second order term reads
\begin{align}\label{eq:V6}
   V_6(\bm r) = \frac{1}{(4\pi \epsilon_0)^2r^6}\sum_{\nu} C_\nu (\theta,\phi) .
\end{align}
The expressions for the coefficients $C_{\nu }$ are given in 
{\it End Matter (Appendix B)}.

We validate the effective potential approach by comparing coupled-channel calculations of collisional properties with single-channel calculations for pairs of molecules colliding in the state $\ket{+}\ket +$. We solve the Schr\"odinger equation by the Manolopoulos algorithm \cite{ManolopoulosA} using the interaction $V_\mathrm{dd}$ and impose capture boundary conditions at the short range $R_{\min}=50a_0$~\cite{Janssen2012bv}, see {\it End Matter (Appendix~A)} for further details. We retrieve the $S$-matrix, from which we calculate the elastic ($k_\mathrm{el}$) and inelastic ($k_\mathrm{inel}$) scattering rate coefficients, as well as the short range loss coefficient $k_\mathrm{short}$ and the complex $s$-wave scattering length $\alpha =a_\mathrm s-ia_\mathrm{Im}$ \cite{MukherjeeShielding}. 
The total loss rate is given by $k_\mathrm{loss}=k_\mathrm{inel}+k_\mathrm{short}$. For the single channel calculations, we use the effective potential $V_\mathrm{eff}$ and calculate the $s$-wave scattering length $a_s$.
\begin{figure}
    \centering
    \includegraphics[width=\linewidth]{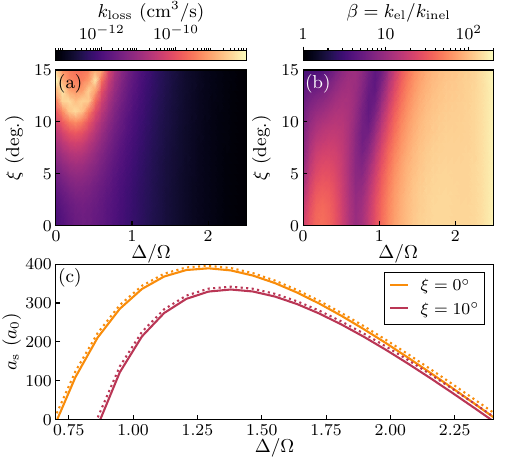}
    \caption{Collisional properties of shielded CaF molecules. (a) Total loss rate coefficient $k_{\mathrm{loss}}$ as a function of relative detuning ($\Delta/\Omega$) and ellipticity ($\xi$) at a fixed electric field of $22.5~\mathrm{kV/cm}$ and $\Omega=60$~MHz, at the energy 10~nK$\times k_B$. (b) Elastic to inelastic collision ratio $\beta$. (c) Real part of the s-wave scattering length from coupled-channel calculations (solid lines) and from single-channel calculations (dotted lines).}
    \label{fig:scatt}
\end{figure}
In Fig.~\ref{fig:scatt}, we show the results for CaF molecules with an electric field $F=$22.5~kV/cm and a Rabi frequency $\Omega=2\pi\times 60$~MHz, which we keep constant throughout this work. Panel (a) shows the total two-body loss rate for the range of detuning considered here. Apparently, for $\Delta/\Omega> 1$, the losses are small even at large values of ellipticity that are normally not possible with other shielding protocols. Panel (b) shows the ratio of elastic-to-inelastic collisions $\beta=k_\mathrm{el}/k_\mathrm{inel}$, which remains large, potentially allowing for efficient evaporative cooling towards quantum degeneracy~\cite{Schindewolf2022}. For example, for typical peak densities $n_\mathrm{peak}\simeq10^{14}~\mathrm{cm}^{-3}$ such as in Fig.~\ref{fig:schem}(b), obtained with $\Delta/\Omega\simeq 2$ and $\xi=0$  the total loss rate is around $10^{-13}$~cm$^3/s$, corresponding to a lifetime of about $0.1$~s. In panel (c) we show the calculated $s$-wave scattering length from the full interaction and from the effective potential, showing quantitative agreement.

In order to employ the effective interaction in mean field calculations, we replace the short range term $V_6$ of $V_\mathrm{eff}$ with a contact term $g\delta(\bm r)$~\cite{XuEffective,ArnoneExploring}, where $g=4 \pi \hbar^2 a_s/M$. We define the dipolar lengths $a_\mathrm{dd}^{(i)} = M C_3^{(i)}/3\hbar^2$ and the corresponding interaction ratios $\varepsilon_\mathrm{dd}^{(i)} = a_\mathrm{dd}^{(i)}/a_\mathrm s$. Under these conventions, the effective interaction potential becomes
\begin{align}
    V'(\bm r) = g \delta(\bm r)+V_3(\bm r)\,.
\end{align}

\indent \prlsection{Model}\label{sec:model}
We simulate the gas of $N$ molecules by using the extended Gross-Pitaevskii equation, which reads
\begin{align}\label{eq:gpe}
    i\hbar\partial_t\psi = \Big( -\frac{\hbar^2}{2M}\nabla^2+V_\mathrm{trap} + V'*|\psi|^2 + g_\mathrm{QF}|\psi|^3\Big)\psi\, ,
\end{align}
where $\psi$ is the order parameter, $V_\mathrm{trap}=(\omega_x^2x^2+\omega_y^2y^2+\omega_z^2z^2)/2$ is the external trapping potential, $g_\mathrm{QF}$ is the beyond-mean-field coefficient \cite{Lima2011}
\begin{align}
    g_\mathrm{QF} = \frac{64 \hbar^2}{3M\sqrt{\pi}} a_s^{5/2} \, \mathrm{Re}\left[\mathcal{K}_5(\varepsilon^{(1)}_\mathrm{dd}, \varepsilon^{(2)}_\mathrm{dd})\right]\,,
\end{align}
and $\mathcal{K}_5$ is the angular integral, given by
\begin{align}
    \mathcal{K}_n(\varepsilon^{(1)}_\mathrm{dd}, \varepsilon^{(2)}_\mathrm{dd}) = &\int_0^{2\pi} \mathrm{d}\phi_k \int_0^1 \mathrm{d}s 
    \big[1 + \varepsilon^{(1)}_\mathrm{dd} \big(3s^2 -1\big) \nonumber \\
    & -3\varepsilon^{(2)}_\mathrm{dd} \cos 2\phi_k (1 - s^2)\big]^{n/2}\,.
    \label{eqn:K5}
\end{align}

\begin{figure}[t]
    \centering
    \includegraphics[width=\linewidth]{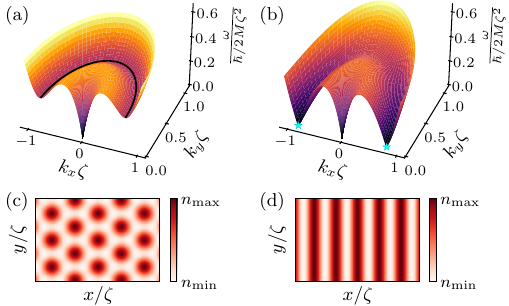}
    \caption{Excitation spectrum of a molecular condensate in planar geometry. We fix the dipolar parameter at $\varepsilon^{(1)}_\mathrm{dd}=1.75$ and  $l_z = 2.88\times\zeta$. (a) and (b) show the spectrum for $\varepsilon^{(2)}_\mathrm{dd}=0$ and $\varepsilon^{(2)}_\mathrm{dd}=0.05$, respectively. The solid line in (a) and the stars in (b) highlight the energy minima. Panels (c) and (d) schematically show the expected groundstate density upon roton softening for (a) and (b), respectively.}
    \label{fig:roton}
\end{figure}

\prlsection{Anisotropy-induced stripe formation} We start by considering a planar infinite system by setting $\omega_x=\omega_y=0$. The ground state wavefunction can then be written as $\psi(\bm r)=\sqrt{n_\mathrm{2D}}\psi_z(z)$. In order to obtain an analytical model, we approximate the wavefunction in the  $z$-direction by a Gaussian, $\psi_z=\exp(-z^2/(4l_z^2))/(2\pi l_z^2)^{1/4}$. The correct value of $l_z$ can be obtained by minimizing the ground state energy for a given trap frequency $\omega_z$. Assuming that the confinement is sufficiently strong, we can integrate out the transverse degree of freedom to obtain an equation for the two-dimensional wavefunction $\psi(x,y)$. To obtain collective excitations, we linearize Eq.~(\ref{eq:gpe}), leading to the corresponding Bogoliubov-de Gennes (BdG) equations \cite{Pethick2008,Schubert2026} that we solve at fixed homogeneous 2D density $n_\mathrm{2D}$. The result for the excitation energy $\hbar\omega$ is the well known expression for the Bogoliubov dispersion 
$$\hbar\omega=\sqrt{\varepsilon_k[\varepsilon_k+2n_{2\rm D}U_{2\rm D}(\bm k)]}~,$$ 
where $\varepsilon_k=\hbar^2k^2/2M$ and
\begin{align}\label{eq:U2D}
    U_\mathrm{2D}(\bm k) &= \frac{3 g_{QF}n_\mathrm{2D}^{1/2}}{ \sqrt 5 l_z^{3/2} 2^{5/4}\pi^{3/4} }+\frac{g}{2} \bigg(\frac{2 \varepsilon^{(1)}_\mathrm{dd}+1}{\sqrt{\pi } l_z} \\
    &-\frac{e^{k^2 l_z^2} \text{erfc}\left(k l_z\right) \left(3 k^2 \varepsilon^{(1)}_\mathrm{dd}+\varepsilon^{(2)}_\mathrm{dd} \left(k_x^2-k_y^2\right)\right)}{k}\bigg)~. \nonumber
\end{align}
In Fig.~\ref{fig:roton} we show the excitation spectrum of a planar gas of molecules, scaling quantities with respect to the healing length $\zeta = \hbar(Mn_\mathrm{2D}g/(\sqrt \pi l_z))^{-1/2}$. For a perfectly circular microwave, the interaction is completely isotropic, and the dispersion relation does not depend on the azimuthal angle. This dispersion exhibits a degenerate set of roton modes with finite momentum $k$. Increasing the dipolar length the minimum softens, leading to the dynamical instability of the unmodulated ground state towards supersolid formation. Such collapse gives rise to a triangular lattice \cite{Lu2015}, which is sketched in Fig.~\ref{fig:roton}(c). At a finite ellipticity, the interaction acquires an angular dependence, distorting the dispersion relation and lifting the rotational degeneracy. In this case, the softening of the roton occurs only at two critical values $\bm k_c=(\pm k_c,0)$, implying that density modulations appear only along the $x$ direction. This results in a stripe phase, as in Fig.~\ref{fig:roton}(d).

\begin{figure}
    \centering
    \includegraphics[width=\linewidth]{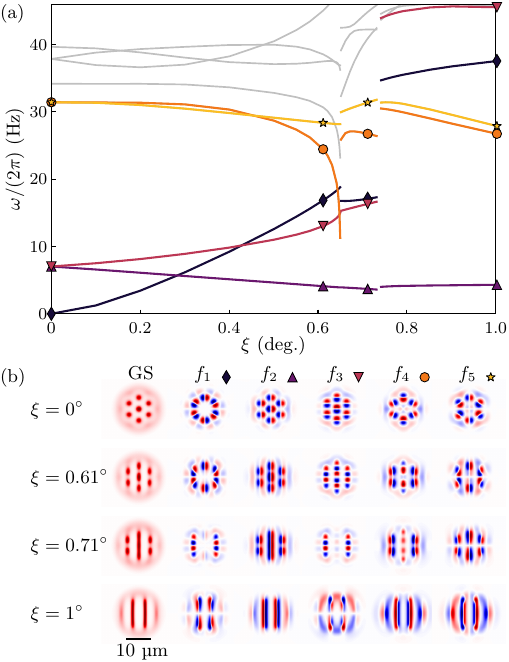}
    \caption{Excitations of a confined planar molecular supersolid of $3\times10^4$~CaF molecules under an electric field $F=22.5$~kV/cm and microwave of Rabi frequency $\Omega=60$~MHz and detuning $\Delta=1.925\Omega$. (a) Excitation spectrum as a function of the ellipticity $\xi$. Colored lines correspond to modes we show in panel (b). (b) Ground state density and low-lying excitation modes $f_i$ in the $z=0$ plane for four representative values of $\xi$. The first column shows the ground state density, other columns show the excitation modes the frequency of which is marked in panel (a).}
    \label{fig:spectrum}
\end{figure}

In the following, we consider a finite-sized system of $3 \times 10^4$ CaF molecules in a harmonic trap with trapping frequencies $(\omega_x,\omega_y,\omega_z)=2\pi\times (50,50,180)$~Hz. We determine the ground state the system by propagating Eq.~\eqref{eq:gpe} in imaginary time and calculate the spectrum of excitations by numerically solving the BdG-equations, which yields discrete frequencies $\omega_i$ and mode functions $f_i$ \cite{Schubert2026b} due to the finite size of the system. The mode functions $f$ are related to the density fluctuations via $\delta n\propto f\psi_0$. In our calculations, we fix the relative detuning $\Delta /\Omega= 1.925$ and increase the ellipticity $\xi$, see Fig.~\ref{fig:spectrum}. In this parameter range, the dipolar length $a_\mathrm{dd}^{(1)}$ is strictly constant at $399 a_0$, the $s$-wave scattering length is roughly constant at $a_s=229a_0$ and the dipolar length $a_{\rm dd}^{(2)}$ is tuned by varying the ellipticity such that it increases linearly from zero to $25a_0$ at $\xi=1^\circ$.

Starting from $\xi = 0$, the system’s ground state consists of seven droplets arranged in a triangular lattice, see Fig.~\ref{fig:schem}(b). Due to the rotational symmetry of the trap and the interaction potential, the lattice structure emerges spontaneously, which results in a zero-energy Goldstone mode $f_1$, marked by black diamonds in Fig.~\ref{fig:spectrum}(a), in addition to the Goldstone mode associated with the spontaneously broken U(1) symmetry. This mode increases in energy once we increase $\xi$. Additionally, we observe two low-lying phonon modes, $f_2$ (purple line) and $f_3$ (red line), which correspond to the lowest lying sound modes (the center-of-mass oscillations) in the $x$ and $y$ directions, respectively. These modes remain finite as a consequence of the transverse confinement. At $\xi = 0$, these modes are degenerate, reflecting the initial symmetry of the Hamiltonian. The degeneracy is lifted upon increasing $\xi$, in good agreement with the prediction of two distinct sound velocities in the $xy$-plane. The same splitting for $\xi > 0$ is also visible for the next two higher sound modes ($f_4$, orange and $f_5$, yellow line).

Upon further increasing the ellipticity, the three droplets along $x = 0$ merge. This marks the onset of the transition into a striped phase,  characterized by the corresponding discontinuity in the excitation spectrum at $\xi = 0.64$. This indicates that the transition is of first order. Interestingly, it is accompanied by a sharp drop in the second-lowest sound mode for $\xi < 0.64$. This mode drives density fluctuations by bringing the two outer droplets at $x = 0$ closer together, directly facilitating the merging of the droplets. Due to the sudden reorganization of the density along $y$, only the lowest- lying sound mode in $x$ is continuous at the phase transition. In good agreement with our model, the stripe phase becomes more pronounced upon further increasing $\xi$, leading to another first-order phase transition at $\xi = 0.73$, beyond which the ground state consists of two well-pronounced stripes, see the groundstate density isosurface Fig.~\ref{fig:schem}(c) for $\xi =1^\circ$. 

Because the interactions are dominated by the dipolar length $a_{\rm dd}^{(1)}$, we define the gas parameter as $\gamma=[a_{\rm dd}^{(1)} ]^3 n_{\rm peak}$. For all values of $\xi$ considered in this work, we find $\gamma < 1.5 \times 10^{-3}$, confirming that the shielding provides a weakly interacting system where mean-field theory is applicable. However, the stripe phase is not unique to small $\gamma$. It also emerges at much larger values of $\gamma$, as demonstrated in double-microwave shielded molecules \cite{Melero2026} and molecules shielded with the shielding protocol considered in this paper \cite{Ciardi2026}.

\indent \prlsection{Conclusion} We derived the effective interaction potential for molecules under combined static electric field and microwave shielding. Through coupled-channel calculations, we show that the potential is reliable and the collisions are mostly elastic, even at large values of the ellipticity, maintaining long lifetimes at collisional energies as low as 10~nK. We investigated the effects of such an interaction on a planar gas of molecules in two ways. First, we derived the excitation spectrum of an infinite gas via a variational model, which predicts stripe formation even at very small ellipticity. Then, we calculated the ground state and excitations of a confined gas, where increasing the ellipticity causes a splitting of each sound mode and a lifting of the zero energy Goldstone mode connected to translational invariance.

 We find that realizing a supersolid in the form of a triangular lattice would require an almost perfectly circular microwave field, which is experimentally challenging. However, this limitation can be overcome by 
orienting the electric field in the $xy$--plane and tuning the orthogonal microwave close to the compensation point, where $a_\mathrm{dd}^{(1)}\simeq0$ such that one recovers a similar dipolar interaction with larger ellipticities. 

Our work shows that extending the recently suggested shielding scheme by Ho {\it et al.}~\cite{HoTuning} to elliptically polarized microwaves may open a new route towards creating novel molecular supersolids with controllable anisotropic interactions. 

\indent\prlsection{Acknowledgments}
This work was financially supported by the Knut and Alice Wallenberg Foundation (Grant No. KAW 2023.0322) and the Swedish Research Council (Grant No. 2022-03654VR). Part of the computations were enabled by resources provided by the National Academic Infrastructure for Supercomputing in Sweden (NAISS), partially funded by the Swedish Research Council through Grant Agreement No. 022-06725VR.

\indent\prlsection{Data availability}
The data supporting the findings of this article are not publicly available, but can be obtained from the authors upon reasonable request.

\bibliography{refs}

\section{End Matter}\label{app:effpot}
\prlsection{Appendix A: Details}
From the one-molecule basis set $\mathcal B$ introduced in the main text, we construct the two-molecule basis $\{\ket\nu\}$. In particular, we highlight the states
\begin{equation}
\label{eq:basis}
\begin{array}{c|c}
|\nu\rangle & E_\nu/\hbar \\ \hline
\multicolumn{1}{l|}{\ket{1}=\ket{+}\ket{+}} & \Omega_{\rm eff} - \Delta \\ 
\multicolumn{1}{l|}{\ket{2}=\ket{+}\ket{\xi_2}} & (\Omega_\mathrm{eff}-3\Delta)/2 \\ 
\multicolumn{1}{l|}{\ket{3}=\ket{+}\ket{-}} & -\Delta \\ 
\multicolumn{1}{l|}{\ket{4}=\ket{-}\ket{\xi_2}} & -(3\Delta+\Omega_\mathrm{eff})/2 \\ 
\multicolumn{1}{l|}{\ket{5}=\ket{-}\ket{-}} & -\Delta-\Omega_\mathrm{eff} \\ 
\multicolumn{1}{l|}{\ket{6}=\ket{0,0}\ket{2,0}} & -\Delta_\mathrm S \\ 
\multicolumn{1}{l|}{\ket{7}=\ket{0,0}\ket{2,1}} & -\Delta_{\mathrm S,2} \\ 
\multicolumn{1}{l|}{\ket{8}=\ket{0,0}\ket{2,-1}} & -\Delta_{\mathrm S,2} \\
\end{array}
\end{equation}
where symmetrization is implicit and $\hbar\Delta_{\mathrm S}=E_{0,0}+E_{2,0}-2E_{1,0}$. Note that in addition to the basis considered in Ref. \cite{HoTuning}, we included the states $\ket 7$ and $\ket 8$, at the energy $-\hbar\Delta_{\mathrm S,2}=E_{0,0}+E_{2,1}-2E_{1,0}$, which contribute significantly to the effective interaction.
In the center of mass frame, two molecules in free space are governed by the following Schr\"odinger equation
\begin{align}\label{eq:sch}
    -\frac{\hbar^2}{2\mu}\nabla^2 \psi_\nu(\bm r) + \sum_{\nu'}V_\mathrm{\nu,\nu'}(\bm r)\psi_{\nu'}(\bm r)= (E-E_\nu) \psi_\nu(\bm r)
\end{align}
where $\mu=M/2$ is the reduced mass and the collision energy is $E=E_1+\frac{\hbar^2k^2}{2\mu}$. The matrix element $V_{\nu,\nu'}$ is given by
\begin{align}
    V_{\nu,\nu'}(\bm r) =  \bra{\nu}V_\mathrm{dd}(\bm r)\ket{\nu'} ~.
\end{align}

We solve Eq.~(\ref{eq:sch}) by means of the log-derivative Johnson algorithm to evaluate the $S$-matrix and the $T$-matrix, $S=1+T$, which we express as $S_{\nu,\ell, m}^{\nu',\ell',m'}$ introducing the quantum numbers of orbital angular momentum $\ell,m$. The states in Eq.~\eqref{eq:basis} are considered explicitly in the calculation, while other states are taken into account by a Van Vleck transformation \cite{VanOn, MukherjeeShielding}. We then evaluate the elastic and inelastic scattering cross sections \cite{MukherjeeShielding} as
\begin{align}
    \sigma_\text{el} &= \frac{2\pi^2}{k^2}\sum_{\ell m \ell' m'} \left|T_{1,\ell,m}^{1,\ell',m'}\right|^2,\\\sigma_\text{in} &= \frac{2\pi^2}{k^2}\sum_{\nu'\neq  1}\sum_{\ell m,\ell'm'}\left|T_{1,\ell,m}^{\nu',\ell',m'}\right|^2,
\end{align}
respectively. The short range loss is calculated from the unitary deficit of the $S$-matrix,
\begin{align}
    \sigma_\mathrm{short} = \frac{2\pi^2}{k^2}\sum_{\ell m}\left(1 - \sum_{\nu'\ell'm'}\left|S_{1,\ell,m}^{\nu',\ell',m'}\right|^2 \right)
\end{align}
while the rate coefficients can be computed from the corresponding cross sections as $k_i = \sigma_i v_\mathrm{rel}$, where $v_\mathrm{rel}=\hbar k/\mu$ is the relative velocity. Finally, the s-wave scattering length is given by
\begin{align}
    \alpha = \lim_{k\to0}\frac{1-S_{100}^{100}(k)}{ik(1+S_{100}^{100}(k))}.
\end{align}
The single-channel Schr\"odinger equation reads 
\begin{align}\label{eq:single}
    -\frac{\hbar^2}{2\mu}\nabla^2\psi(\bm r) +V_\mathrm{eff}(\bm r)\psi( \bm r) = \frac{\hbar^2 k^2}{2\mu} \psi(\bm r).
\end{align}
In this case, the (elastic) cross section is
\begin{align}
    \sigma_\text{eff}=\frac{2\pi^2}{k^2}\sum_{\ell m, \ell' m' }|T_{\ell m}^{\ell'm'}|^2
\end{align}
and the s-wave scattering length is 
\begin{align}
    a_\mathrm{s} = \lim_{k\to0}\frac{1-S_{00}^{00}(k)}{ik(1+S_{00}^{00}(k))}.
\end{align}

\prlsection{Appendix B: Coefficients}
The first order term in the effective interaction is given by
\begin{align}
    V_3(\bm r) = \bra{1}V_\mathrm{dd}(\bm r)\ket 1 .
\end{align}
A simple calculation gives $V_3$ as in Eq.~\eqref{eq:V3}. The second order term is instead given by Eq.~\ref{eq:V6}, where 
\begin{align}
    V_6(\mathbf r) = \sum_{\nu\neq1}\frac{|\bra 1 V_\mathrm{dd}(\bm r) \ket \nu|^2}{E_1-E_\nu}~.
\end{align}
The matrix elements are given by $|\bra 1 V_\mathrm{dd} \ket \nu|^2=C_\nu/(4\pi\epsilon_0 r^3)^2$, with angle-dependent coefficients
\begin{align}
    C_2 &= \frac{9}{2} (d^{(+)}_{10,11})^4 u^4 v^2 \sin\theta^4 (1 - \cos^2 2 \phi \sin^22 \xi) \nonumber\\
    C_3 &= \frac{1}{8} u^2 v^2 (((d^{(+)}_{10,11})^2 (u^2 - v^2) - 2 (d^{(z)}_{11,11} - d^{(z)}_{10,10})\nonumber\\ &\times (d^{(z)}_{10,10} u^2 + d^{(z)}_{11,11} v^2))(1 + 3 \cos 2 \theta) \nonumber\\ &+ 6 (d^{(+)}_{10,11})^2 (u^2 - v^2) \cos2 \phi \sin^2\theta \sin2 \xi)^2\nonumber\\
    C_4 &= \frac{9}{2} (d^{(+)}_{10,11})^4 u^4 v^2 \sin\theta^4 (1 - \cos^22 \phi \sin^2 2\xi) \nonumber\\
    C_5 &= \frac{1}{4} u^4 v^4 (((d^{(+)}_{10,11})^2 + (d^{(z)}_{11,11} - d^{(z)}_{10,10})^2) (1 + 3 \cos 2\theta) \nonumber\\
    &+ 6 (d^{(+)}_{10,11})^2 \cos 2\phi \sin^2\theta \sin 2 \xi)^2 \nonumber\\
    C_6 &= \frac{1}{2} (d^{(z)}_{00,10})^2 (d^{(z)}_{10,20})^2 u^4 (1 + 3 \cos 2 \theta)^2 \nonumber\\
    C_7 &= 9 (d^{(+)}_{10,21})^2 (d^{(z)}_{00,10})^2 u^4 \cos^2\theta \sin^2\theta \nonumber\\
    C_8 &= 9 (d^{(+)}_{10,21})^2 (d^{(z)}_{00,10})^2 u^4 \cos^2\theta \sin^2\theta
\end{align}


\end{document}